\documentclass[aps,prl,reprint,showpacs]{revtex4-1}

\usepackage{amsmath}    

\usepackage{graphicx}   
\usepackage{verbatim}   
\usepackage{color}      
\usepackage{subfig}     
\usepackage{hyperref}   
\usepackage{bm}         
\usepackage{amssymb}    

\begin{document}

\title{Rare Events, Extremely Rare Events and \\  Fluctuations in a Thermodynamic System}
\author{P.J. Malsom}
\author{F.J. Pinski}
\affiliation{Department of Physics, University of Cincinnati, Cincinnati, Ohio 45221, USA}

\date{\today}

\begin{abstract}

In this paper, we follow in the footsteps of Onsager and Machlup (OM) and consider diffusion-like paths that are explored by a particle moving via a conservative force while being in thermal equilibrium with its surroundings. 
Instead of considering diffusion (Brownian dynamics), we use a Metropolis algorithm to derive an OM-like functional.
Through the lens of the Metropolis algorithm, we are able to elucidate the errors made when using a nonzero time increment.

Of particular interest are transition paths that transverse an energy barrier that is large (but not too large) compared to the typical thermal energy.
These transitions have probabilities that are only small and yet not so small as to be considered a violation of thermodynamics.
As such, we turn our attention to the "double-ended" problem, where the OM functional can be interpreted as a "thermodynamic" action and employed to sample paths that start and end at predetermined points.
We find that sampling the continuous-time limit of the OM functional, i.e., the Ito-Girsanov form, produces unphysical paths due to thermodynamic inconsistency arising from the singular nature of the limit. 
These unphysical effects are due to the correlation between the particles position and the random fluctuating noise that is introduced in the limiting process.
Such a correlation is a direct consequence of the form of the Ito-Girsanov functional, does not originate from numerical inaccuracies, and is a violation of thermodynamics as embodied by Seifert's Integral Fluctuation Theorem.

\end{abstract}
\maketitle

\section{\label{sec:level0}{Introduction}}

The aim of this paper is much the same as the 1953 article{\cite{Onsager:1953}} of Onsager and Machlup, namely, to derive the probability of a succession of states of a spontaneously fluctuating thermodynamic system.
Their expression for this (path) probability has become known as the Onsager-Machlup (OM) functional and has so far withstood the test of time.
One commentary even stated that the results of this 1953 paper are "incapable of improvement either in form or in their mode of derivation."{\cite{McKean:98}}

Here we follow the spirit of the OM work, and try to fully understand its limitations while exploring the continuous-time limit.
The derivation of the OM functional that we report here, has its roots in a different 1953 article by Metropolis \textit{et al}.{\cite{metropolis:1953}}
While the OM function can be derived by considering diffusion (in terms of Brownian dynamics), one can use the Metropolis algorithm to correct for any errors due to the nonzero time step.
The strategy here is to use a variant, the Hybrid Monte Carlo  (HMC) algorithm{\cite{Duane:1987}}.
Using this approach, we derive an analogous OM functional and show how the functional is consistent with thermodynamics, that is, with the underlying Boltzmann distribution.

The OM functionals used here act as a thermodynamic action{\cite{graham1977}} for generating paths that are constrained at both ends.
One way to proceed is to take the continuous time limit of the functional; to use Ito calculus and the Girsanov theorem{\cite{DurrBach:1978}}.
We have used this limiting functional to generate an ensemble of paths and found them to be unphysical.
We show this unphysical nature  originates in the form of the Ito-Girsanov measure, which violates Seifert's Integral fluctuation theorem{\cite{IFT:2005}}.

\section{\label{sec:level2}{Brownian Dynamics}}

Throughout this paper, we will consider a particle in contact with a heat reservoir at a temperature $\epsilon$.
It is moving under the influence of a potential ${\mathbb{V}}(x)$ with the force being $F(x)= -{\mathbb{V}}'(x)$.
Note that although the equations are written for the one-dimensional case for clarity, the formalism can easily be extended to higher dimensions and for a collection of particles.

 The equation of motion for Brownian dynamics is given by
\begin{equation}
    dx = F(x) \,dt + \sqrt{2 \epsilon} \, dW_t
    \label{SDE}
\end{equation}
 where $dW_t$ is the standard Weiner process that represents the (uncorrelated) Gaussian noise.
Using a discrete time step, $\Delta t$, one typically uses the Euler-Maruyama algorithm{\cite{Maruyama1955}} as an approximate method for propagating the position as a function of time.
Namely,
\begin{equation*}
    x_{i+1} = x_i + F(x_i)\, \Delta t + \sqrt{2 \, \epsilon \, \Delta t \,} \, \xi_i
\end{equation*}
where $\xi$ is a Gaussian random variate with mean zero and unit variance.
Successive application (N times) of this equation produces a sequence of positions $\{x_i\}$ which is called a path.
Onsager and Machlup{\cite{Onsager:1953}} used the underlying thermal fluctuations to write the Gaussian probability, $\mathbb{P}_p \propto \Pi_i \exp{\!( - \frac{1}{2}  \xi_i^2 )} $, of the path in terms of the path variables themselves, namely,
\begin{equation}
    - \log{ \mathbb{P}_p} =
    \frac{\Delta t}{ 2\, \epsilon}\, \sum_{i=1}^N
    \frac{1}{2} \Big( \, \frac{ x_{i+1}-x_i}{\Delta t} - F(x_i)  \Big)^2 \,.
    \label{ProbOM}
\end{equation}
This equation is commonly called the OM functional.
In the continuous-time limit, using Ito calculus and the Girsanov theorem, the Radon-Nikodym derivative is used to express the change in the measure:
\begin{equation}
    \frac{d \mathbb{P}_{p}}{d \mathbb{Q}_{p}} =  \exp{  \!  \Bigg( \! \!-\frac{1}{2 \, \epsilon}
    \Big(  \mathbb{V}(x_T)  - \mathbb{V}(x_0)  + \!  \int_0^T \! \! \!\!  dt \ G(x_t)   \Big) \Bigg)  }
    \label{ProbIto}
\end{equation}
where $T$ is the duration of the path, $\mathbb{Q}$ is the measure associated with free Brownian dynamics, and the function $G(x)$ is defined as  $G(x)=\frac{1}{2} F(x) \cdot F(x)  - \epsilon \,  \mathbb{V}''(x)$.

\section{\label{sec:level2a}{Some Results}}

One of the uses of the OM functional is to incorporate it into a scheme to sample paths that are constrained at both ends.
The aim is to efficiently generate an ensemble of paths that include a transition over an energy barrier.
When the barrier is large compared to the typical thermal energy, the transition is a rare event. 
Such barrier hopping is consistent with thermodynamics with the noise reflecting the fluctuating random effects that are independent of the particle's position. This contrasts to what we designate as extremely rare events where an occurrence would seem to violate thermodynamics, for example when all the molecules in a room migrate to one corner. 
Here we only consider the former.

Conventional forward time integration is not particularly efficient, especially for small $\epsilon$, since the hopping rate is small.
We incorporate the Ito-Girsanov formula into a Monte Carlo method and sample the measure which, after some initialization, is used to generate paths that reflect the probability distribution.
We use a recently devised path-space sampling method{\cite{Beskos}}.

\begin{figure}[t] 
    \includegraphics[scale=.5]{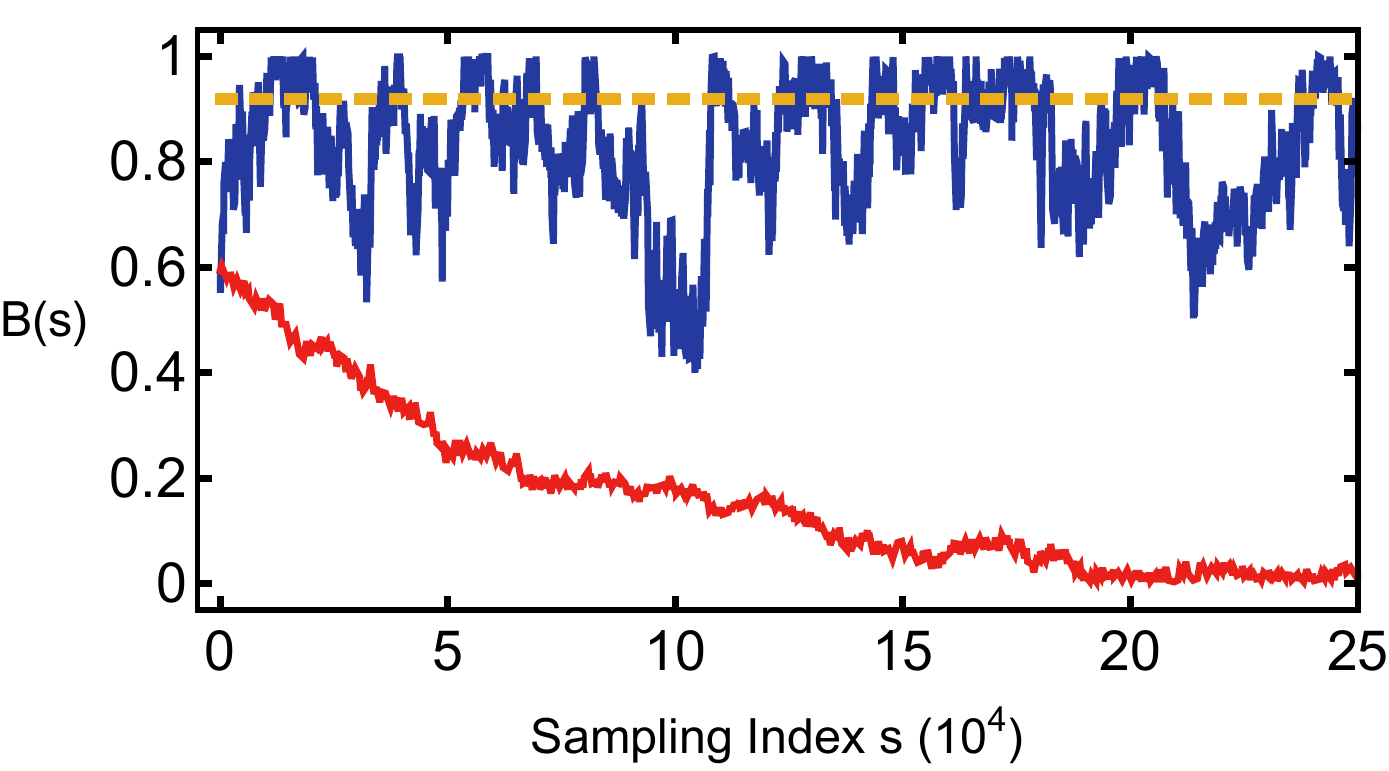}
    \caption{
        The fraction of the path in the broad well is shown as a function of an arbitrary sampling index, $s$.
        The results from both methods start with the same input path where $B(0) \approx 0.6$, have a path length of $T = N \, \Delta t = 150$, and a time step along the path of $\Delta t = 0.005$.
        The lower (red) curve corresponds to the results using the Ito-Girsanov form, while the upper (blue) curve are the results of using the mid-point functional.
        The (orange) dotted line corresponds to the equilibrium value.
    }
    \label{BroadFractionFig}
\end{figure}

The quandary is that for a very simple one-dimensional example, the generated paths  quickly become unphysical.
Long paths, generated with small time steps, are expected to be consistent with equilibrium thermodynamics.
The positions along the path should be distributed according to the Boltzmann probability, $\exp{  (  -\mathbb{V }/ \epsilon ) }$.
As an example, consider the potential
\begin{equation}
    {\mathbb{V}}(x)  = 2^{-26} (    8 -5 x   )^8 \, ( 2 + 5x )^2
    \label{ThinBroadPot}
\end{equation}
which has two degenerate wells with a barrier of unity at the origin.
A narrow (quadratic) well is on the left and the wide well is on the right.
At any nonzero temperature, the particle will spend more time on the right due to the entropic considerations inherent in the Boltzmann distribution.
We used the Ito-Girasanov formulas with the method developed by Beskos et al{\cite{Beskos}}, at a temperature $\epsilon =0.25$, to generate a sequence of these paths.
We use the Heaviside function $\Theta$ to define $B(s)$ as the fraction of the path that is contained in the broad well, namely
\begin{equation*}
    B(s) = \frac{1}{T} \int_0^T dt  \ \Theta(x_t^{(s)})  \approx \frac{1}{N}  \sum_i \Theta(x_i^{(s)})
\end{equation*}
where the sampling index is denoted as $s$, and the corresponding path is $\{\,x_i^{(s)}\}.$
In Figure $\ref{BroadFractionFig}$, we plot $B(s)$ using the described procedure (red curve) and compare it to one using a second method described in a later section.
As can be seen from the figure, the paths quickly become unphysical in that the particle spends the vast majority of the time in the left, thin well.
Such paths are inconsistent with the equilibrium thermodynamical distribution.
Thus the purpose of this article is to explain the origins of these unphysical results and to describe which functionals can be used to sample physical paths with double-ended boundary conditions.

\section{\label{sec:level3}{A New perspective}}

Given the difficulty described above, we turn to an alternate derivation of the OM functional.
In doing so, we expose where the errors are hidden in the traditional view of Brownian dynamics.
Instead of basing a path-probability functional on Brownian dynamics, we will use the Metropolis-based, Hybrid Monte Carlo (HMC) method{\cite{Duane:1987}}.
In the incarnation used here, velocities are chosen from the temperature-appropriate Maxwell-Boltzmann (Gaussian) distribution.
The deterministic Hamilton's equations are integrated over a single time step $h$.
The Metropolis acceptance step is determined by the change of the total energy that occurred during the deterministic integration.
This procedure is sometimes called Metropolis Adjusted Langevin Algorithm (MALA) and Smart Monte Carlo{\cite{Rossky:1978}}.
Thus in an appropriate limit, the procedure has the random-walk properties of Brownian motion{\cite{nealmcmc:2010}}.
Note that if the integration were exact, all moves would be accepted.

The details are as follows.
The configuration or position is $x_i$ at time $t_i$.
The mass matrix is chosen to be the identity matrix.
The velocities at the start of the interval are drawn from the Gaussian distribution with mean zero and variance set equal to the temperature.
Then one integrates (approximately) Hamilton's equations over a time $h$.
First we consider the leapfrog (or velocity-Verlet) scheme{\cite{leapfrog}}.
The velocities at the beginning of the time interval are defined to be $v_{0,i}= \sqrt{\epsilon} \ \xi_{i}$.
Then $x_{i+1}$ is formed by
\begin{equation}
    x_{i+1} = x_{i} + h\, v_{0,i} + \frac{h^2}{2} F(x_i) \,  .
    \label{LeapFrogPos}
\end{equation}
The velocity at the end of the interval  $v_{f,i}$ is given by
\begin{equation}
    v_{f,i} =  v_{0,i} + \frac{h}{2} \,  \Big(  F\big( x_i \big)  +F\big( x_{i+1} \big)  \, \Big) .
    \label{LeapFrogVel}
\end{equation}
The error in the energy due to using this approximate quadrature is
\begin{align}
\begin{split}
   & \Delta E_i  =
        \frac{1}{2} \, (x_i - x_{i-1} ) \cdot  \Big(  F(x_{i})+ F(x_{i-1})\Big)   \\ &
        + \mathbb{V}(x_i) - \mathbb{V}(x_{i-1})
    +\frac{h^2}{8}\, \Big(  \Big|  F(x_i)  \Big|^2-  \Big| F(x_{i-1})  \Big|^2\Big)    .
    \label{LeapFrogE}
\end{split}
\end{align}
The proposed position $x_{i+1}$ should be accepted or rejected using  the Metropolis-Hastings{\cite{HASTINGs:1970}} test: If $ \exp{\! (- \Delta E_i / \epsilon)} > \eta $, the proposed move $x_{i+1}$ is accepted, otherwise  $x_i$ is reused. 
Here $\eta \in   \mathbb{R}$  is a random number uniformly chosen on the unit interval, $0<\eta<1$.
The acceptance probability is given by the Metropolis function $\mathbb{M} (\Delta E_i / \epsilon)  = \textrm{Min}(1, \exp{\!(-\Delta E_i / \epsilon)} )$.
Note that without rejections, detailed balance is broken.

\section{\label{sec:level5}{Connection to Brownian Dynamics}}

The paths (sequence of configurations) generated by the HMC method are similar to those generated by iterating the stochastic differential equation (SDE) of Brownian motion.
This can be seen by identifying $\Delta t= h^2/2$ in equation $\ref{LeapFrogPos}$.
This identification is consistent with the recent work of Spivak et al{\cite{Sivak:2013}}.
However, to fully recover the Brownian SDE, the rejection step of the Monte-Carlo algorithm must be ignored.
This leads to a sampling error that will be small if the step size is small.

Thus the two methods sample different measures.
The HMC method, with rejections, is designed to sample the target Boltzmann distribution.
Using Brownian dynamics, without rejections, one samples a "near-by" distribution.
The measures can be (in some sense) close to each other, but only if the forward integration of the SDE corresponds to a HMC method with a low rejection rate.
The rejection rate is governed by the energy error (equation $\ref{LeapFrogE}$).


Before continuing, it is instructive to re-express the OM functional for a path.
The path is given by a sequence of configurations, ${x_i}$.
The path probability is given by $\mathbb{P}_{path} \propto \exp{ \!( - \Phi/ 2 \epsilon)} $ where the functional $\Phi $ (up to an unimportant constant) is expressed as
\begin{align}
\begin{split}
  \frac{ \Phi } {  \Delta t } =& \sum_{i=1}^N   \Bigg(  \frac{1}{2} \Big| \, \frac {\Delta x_{i} } {\Delta t}   \,\Big|^2 
        +\frac{1}{4} \Big| \,F(x_{i}) \,\Big|^2 +\frac{1}{4} \Big| \,F(x_{i-1}) \,\Big|^2  \\
    & \qquad 
        +\frac { \Delta x_{i} } {2 \,\Delta t}  \cdot \Delta F(x_{i})  + \frac{ \Delta \mathbb{V}(x_i) }{\Delta t} - \frac{ \Delta E_i}{ \Delta t } \Bigg) 
    \label{LeapFrogPhi}
\end{split}
\end{align}
where the $\Delta$'s refer to changes as the position changes from $x_{i-1}$ to $x_i$ and $\Delta E_i$ is defined in equation $\ref{LeapFrogE}$.
The first term in the functional $\Phi$ becomes the free Brownian measure (in the appropriate limit), and the fourth becomes  $-\epsilon \,\mathbb{V}''$ when the quadratic variation ($\Delta x^2 \approx 2 \, \epsilon \, \Delta t$) is satisfied.
The difference in potential energy telescopes, and the functional includes all components of the continuous time OM functional (equation $\ref{ProbIto}$).
Below we  show the importance of the energy error terms, $\Delta E_i$.
It is well-known that the leapfrog integrator does not give accurate paths{{\cite{hoover2012}}.
This can be seen here as well, since the error in the path is of order one due to this term, and the energy error does not vanish even for a quadratic potential (an Ornstein-Uhlenbeck (OU) process{\cite{OU:1930}}).

It is important to recognize that rejections must be ignored to map the HMC method onto Brownian dynamics.
Since the energy errors, in general, do not vanish, detailed balance is broken in the Metropolis procedure.
The damage to the sampling ensemble is hard to quantify.
In the Brownian picture, the energy-error terms leads  to spurious entropy production as time-reversal invariance is broken.
As the time step decreases, so do these errors.

 Note that Seifert's Integral Fluctuation Theorem (IFT){\cite{IFT:2005}} states that all paths (of the same duration) are equally probable.
This is seen in equation \ref{ProbOM} by noting that $\log{\mathbb{P}_p}$ is simply proportional to the variance of the Gaussian fluctuations and thus is independent of the force, $F$.
We see this in the HMC algorithm, where the choice of the velocities is independent of the particle position and is the source of the stochastic variation. 
The noise can only be a function of the heat reservoir if one is to recover the thermodynamic (Boltzmann) distribution.
When using a functional to generate double-ended paths, any functional that includes a correlation between the positions and the velocities will produce paths that will be inconsistent with the Boltzmann distribution.
Only in extremely rare instances will the noise be systematically correlated with the particle position.
Such instances are those events that are so improbable that they can be viewed as a violation of thermodynamics.

In particular, while inspecting  the Ito-Girsanov measure, we note that some paths have higher probability than others.
This violates Seifert's Integral fluctuation theorem{\cite{IFT:2005}}.
As one explores the paths distributed according to this measure, the sampling procedure will have a tendency to move to the region where the measure indicates it to be more probable. 
In those regions, the noise is highly correlated with the particle positions.
Thus by its very form, the Ito-Girsanov OM functional must produce an ensemble of paths that will not be consistent with the Boltzmann distribution.

\section{\label{sec:level10}{The Midpoint Integrator}}

We can produce better functionals by improving the accuracy of the integrator.
A first attempt is to use a midpoint method, which is related to Stratonovich integration.
During the time $h$, the error in the energy is given by
\begin{equation}
    \Delta E_i  = \mathbb{V}(x_i) - \mathbb{V}(x_{i-1}) +
        (x_i - x_{i-1} ) \cdot F( \overline{X}_i)
    \label{MidPtE}
\end{equation}
with  $ \overline{X}_i = (x_{i+1} +x_i)/2$.
The modified OM functional,  can be written in terms of the above and we get

\begin{align}
\begin{split}
    \frac{ \Phi_m} {   \Delta t } & =  \sum_{i=1}^N   \Big( \  \frac{1}{2} \Big|  \frac {\Delta x_{i} } {\Delta t}   \,\Big|^2
        +\frac{1}{2} \Big|  F( \overline{X}_i)    \,\Big|^2  \\ & - \frac{2 \epsilon}{ \Delta t} \log{\! \Big(
        1 + \frac{1}{2} \Delta t  \,   { \mathbb{V}}''( \overline{X}_i)    \ \Big) } +    \Delta \mathbb{V}_i - \Delta E_i  
      \Bigg) 
    \label{MidPtPhi}
\end{split}
\end{align}
where the log term comes from the Jacobian that results from converting the Gaussian distribution of the variable $\xi$ to the distribution of $x$.
The advantage of $\Phi_m$ over $\Phi$ is that the energy error vanishes for an OU process.
For nonlinear forces the $\Delta E$ terms do not vanish and may be significant in that it leads to spurious entropy production.
These terms break time-reversal invariance or in the Monte-Carlo picture, they break detailed balance.
The paths generated using the midpoint integrator will of be higher fidelity than those generate by the HMC with the leapfrog integrator, because the energy error and the spurious entropy production are smaller.

With the midpoint method, we use the functional described in equation $\ref{MidPtPhi}$, with the path sampling scheme based on an HMC implementation using the "Implicit Algorithm" in Beskos \textit{et al.}{\cite{Beskos2008}} As shown in Figure $\ref{BroadFractionFig}$, we find that sampled paths are very similar to those based on integrating forward in time as described in equation $\ref{LeapFrogPhi}$.
 In contrast to the results when using  the Ito-Girsanov functional   the value of $B$ fluctuates around a value of $0.8$, about $10\%$ smaller than its equilibrium value.
Compared to a value close to zero found previously, this is a dramatic improvement.

\section{\label{sec:level18}{The Zero-Temperature Limit}}

At zero temperature, one can consider the path that starts in one well, ends in an other, and transverses the barrier.
Large Deviation theory{\cite{FreidWentz84}} coupled with the Ito-Girsanov functional provides a solution that corresponds to a heteroclinic orbit in an effective potential defined by $V_{eff}(x)=- |F(x)|^2 /2$.
Such orbits take an infinite time.
However, one can find approximate minimizers to the corresponding Euler-Lagrange equations.
These approximate minimizers are characterized by long periods of time spent at the critical points (including the maximum, i.e. the barrier).

At zero temperature such large-deviation solutions of the Ito-Girsanov functional are not physical{\cite{gamma:2012}}.
When the temperature of the reservoir is zero, fluctuations vanish.
Thus this limit is also singular.
Looking at sequences of paths, using a nonzero time step with monotonically decreasing temperature, we find that the original OM functional ignores the energy errors, as expressed in equation $\ref{LeapFrogE}$.
At very low temperatures, the Metropolis-Hastings rejection rate is very large.
These errors dramatically change the acceptance probability as the temperature approaches zero.
Thus the fidelity of the path becomes increasingly compromised as temperature is decreased.
This is a direct consequence of the uncontrolled approximations inherent in Brownian dynamics.
Thus as the temperature monotonically decreases, the large-deviation solution can not be the limit of the sequence of paths generated by equation \ref{ProbIto} if one rejects states according to the Metropolis-Hastings criterion.

\section{\label{sec:level20}{Discussion and Conclusion}}
The Onsager-Machlup functional is based on Brownian dynamics and provides a good starting point to understand the way a particle moves in response to thermodynamic fluctuations.
However, it contains uncontrolled approximations.
Here we used a novel approach based on the Metropolis Monte Carlo method to elucidate the errors and to show how the underlying measure will stray away from the correct thermodynamical one as either the continuous-time limit or the zero-temperature limit is taken.

The consequences of the singular limit impacts a wide body of work.
The theory for entropy production in nonequilibrium thermodynamics is just one of them.
As an example see the work of Speck et al{\cite{speck2012}}.
For probing the folding of proteins, the recent work of Fujisaki, et al{\cite{ fujisaki2010}} suffered from using the unphysical form of the Onsager-Machlup function, while the older work of Eastman et al{\cite{eastmandoniach2001}} could be improved by using the functional based on the midpoint quadrature as this would decrease the spurious entropy production.
Other works{\cite{facc2006,eric2004}} have to be reevaluated in light of this insight.

The form of the OM functional has been explored before{\cite{adib2008,zuck2000,hunt1981}}.
It is now clear why the issue of the form of the OM measure remained controversial for such a long time.
Without an understanding of the limiting process, the answer remained hidden.
In this paper, we described diffusive motion (Brownian dynamics) in terms of a Monte Carlo process.
This allowed us to clearly express the OM functional for small, nonzero time steps and to explicitly state the errors made in such a representation.
Through the lens provided by the Monte Carlo procedure, we identified additional terms that contribute to the path probability. 
Thus we were able to uncover that the singular nature of the continuous-time limit of the OM measure lies in its very form.

We end with a conjecture.  
How do we interpret the meaning of the Ito-Girsanov change of measure (equation \ref{ProbIto})? 
Consider the set of all paths, $\Pi_F$, generated from solving equation \ref{SDE}, the index indicating the force. 
We speculate that in the continuous-time limit the sets $\Pi_F$ and $\Pi_{F=0}$ are disjoint. 
Then equation \ref{ProbIto} simply gives an indication of the deviation of a free-Brownian path from a path generated when using a nonzero force.
Thus the Ito-Girsanov change of measure is not related to the probability of the path.

\section{\label{sec:level22}{Acknowlegdments}}
We wish to especially thank Robin Ball, Andrew Stuart, Hendrik Weber, Gideon Simpson, and Florian Theil for many lengthy conversations.


%

\end{document}